# Development of charge-exchange injection at the Novosibirsk Institute of Nuclear Physics and around the World

Vadim Dudnikov

*Muons, Inc. 552 N. Batavia Ave. Batavia, IL, 60510, USA*

**Abstract.** The study of charge-exchange injection of protons into accelerators started in 1960 at the Institute of Nuclear Physics of the Siberian Branch of Russian Academy of Science, as proposed by G. I. Budker in connection with the development of the program of the VAPP-4 proton-antiproton collider. Since the purpose was the accumulation of beams with a maximum intensity, and the record intensity of the H- ion beams received by that time was only 70 µA, an important part of the program was the development of methods to produce intense beams of negative hydrogen ions. Charge-exchange injection of particles into accelerators and storage rings is an important application of the charge-exchange technology. Development of charge exchange injection with compensation of ionization energy loss by RF acceleration voltage up to the space charge limit is presented. Accumulation of coasting beam with space charge compensation with weak focusing and strong focusing is described. Accumulation of circulating beam in a storage ring with intensity above space charge limit is presented. Observation, explanation and damping of e-p instability (electron cloud effect) is discussed. Recent developments of charge exchange injection are discussed. Laser ionization of H- in charge exchange injection is reviewed.

## INTRODUCTION

The charge-exchange injection method was discussed by Alvarez in 1951 [1]. However, at that time, the level of development of methods of negative ion production was so low that one could only talk about studying orbits in a stationary magnetic field without the hope of accumulating beams with appreciable intensity. Later, the attractiveness of the charge-exchange injection of protons into accelerators was independently noted by a number of authors [2]. A dedicated study of the problems associated with charge-exchange injection of protons into accelerators was begun in 1960 at the Institute of Nuclear Physics of the SB RAS, on the proposal of G.I. Budker [3] in connection with the development of the program of the VAPP-4 proton-antiproton collider.

Since the purpose was the accumulation of beams with a maximum intensity, and the record intensity of the H- ion beams received by that time was only 70 µA, an important part of the program being implemented was the development of methods for production of intense beams of negative hydrogen ions. Charge-exchange injection of particles into accelerators and storage rings is an important application of the charge-exchange technology [4]. The schematic of charge-exchange injection is shown in Fig. 1a. A beam of accelerated ions of H- or multiply charged ions is introduced into the orbit. They lose electrons in a thin foil and move along an equilibrium orbit, re-crossing the foil that is relatively transparent to stripped ions.

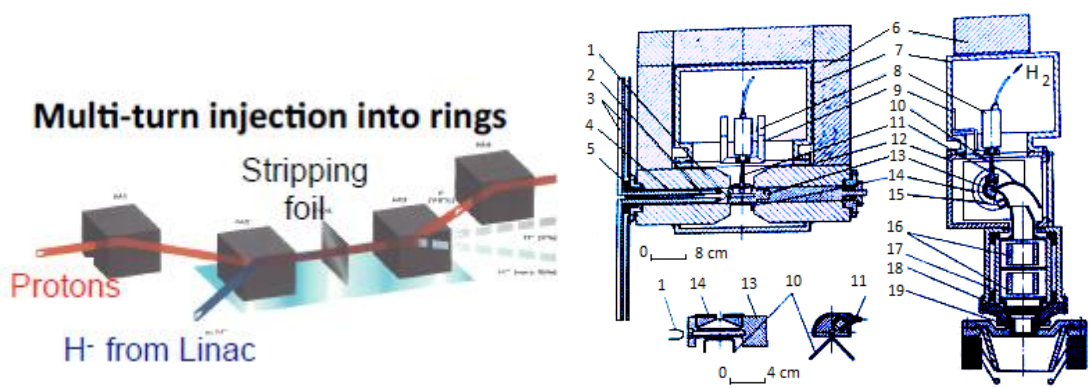

**FIGURE 1 a**. Schematic of charge exchange injection method [7]. b- schematic of an Ehlers type negative ion source in a high voltage terminal Van De Graaf accelerator [7].

Many concrete methods other than charge-exchange for accomplishing the capture of accelerated particles into stationary orbits in the magnetic field of accelerators and storage rings have been worked out: one-turn injection with an inflector, injection into a growing field, spiral accumulation, a method of phase shifting, etc. These traditional methods of injection allow the accumulation of particles into the accelerators for up to hundreds of circulation turns.

In these cases, newly injected particles are placed in the region of the phase space of the system not yet occupied by the previous particles, so that the brightness of the accumulated beam cannot be higher than the brightness of the injected beam. The developed injection technology allows filling, in accordance with the accelerator ring phase space acceptance, at low magnetic field strength. However, the brightness of the available proton beams is not enough to completely fill the acceptance of the storage rings, which have a significantly higher space charge limit. The charge-exchange technology makes it possible to implement an "injector" located on an equilibrium orbit, "transparent" for irreversibly trapped particles. Such an "injector" is the charge exchange target, which "generates" protons. The transformed particles H-, Ho, $H_2+$, $H_3+$, accelerated to the required energy, are directed to the target in such a way that the protons (ions) formed in it move along the necessary trajectories.

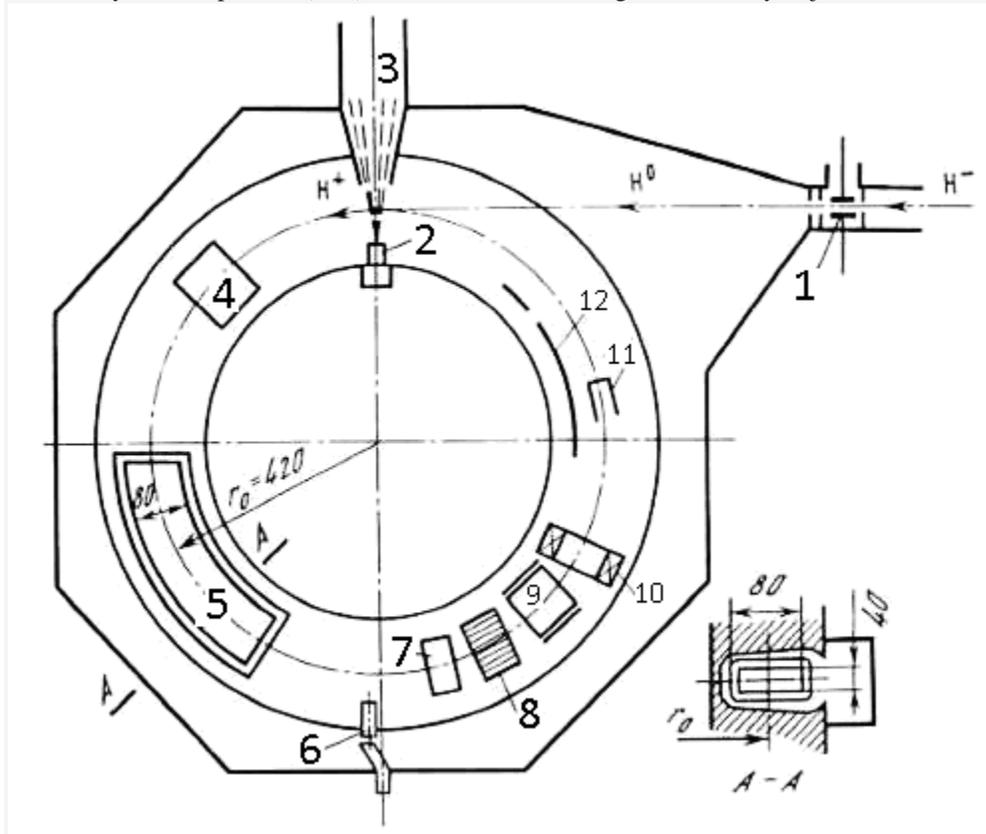

**FIGURE 2**. Diagram of the installation for the study of charge-exchange injection [7].
1-first stripping target, 2-nozzle of supersonic jet, 3- jet receiver, 4-ring pick-up electrode, 5- acceleration drift tube, 6-light collimator of luminescent beam profile monitor, 7- ionization beam intensity meter, 8- ionization profilometer, 9 - pickup-position monitor, 10-Rogowski belt, 11-Faraday cylinder, 12-deflector for suppressing e-p instability.

At the end of the injection, the target can be "removed" and thereby completely eliminate its effect on particle motion. It is essential that in this case new portions of particles fall into the region of phase space already filled with trapped particles, so that the brightness of the accumulated beam can exceed the brightness of the injected beam by orders of magnitude. In doing so, it is possible to overcome the restrictions imposed by the Liouville theorem.

## CHARGE-EXCHANGE INJECTION REALIZATION

Charge-exchange injection of protons into the storage ring was realized experimentally in 1964 [5]. Then it was possible to increase the intensity of the proton beam accumulated on the storage ring by the charge-exchange method up to the space charge limit [6,7,8]. For H- production an Ehlers type negative ion source was used with $LaB_6$ cathode, shown in Fig. 1b [9]. It can produce 1 mA of H-, letter increased to 4 mA. The schematic of the INP SBRAS storage ring for studying charge-exchange injection is shown in Fig. 2. In these experiments, a beam of $H^o$ atoms produced by conversion on a gas target from H- ions, accelerated to an energy of 1 MeV was injected to the charge-exchange target. The efficiency of the H- to-$H^o$ conversion is high, as shown in Fig. 3 [10].

As a stripping target, a supersonic jet of hydrogen was used, which was switched on for the time of injection. The experiments confirmed the initial assumptions

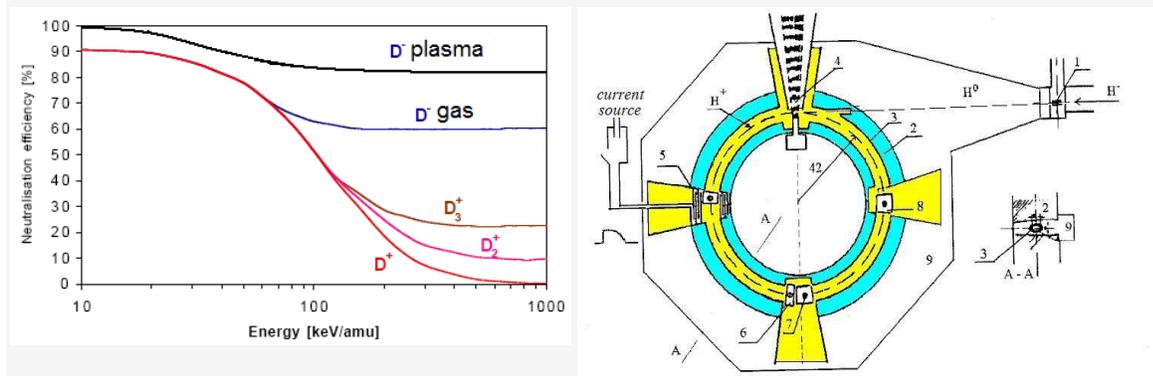

**FIGURE 3**. Generalized data on the achievable efficiency of conversion of hydrogen ion beams to the energy of fast-atom beams at various energies of the atoms obtained [10].

**FIGURE 4**. A storage ring with betatron compensation of ionization energy losses [7].
1-first stripping target, 2-pole magnet, 3-donuts, 4- second stripping target, 5- labyrinth, 6-ring pick-up, 7-ionization current meter, 8 ionization profile monitor, 9 vacuum chamber, ion-electron collector.

The accumulation with ionization losses by high-frequency voltage compensation, the capture efficiency during 2000 revolutions was 75% in accordance with the separatrix area, and when injected during 4000 revolutions the efficiency decreased by only 20%. Up to 300 mA of circulated proton beam was accumulated, limited by the longitudinal space charge limit. The longitudinal space charge served as an additional acceleration RF voltage, increasing the radial synchrotron motion acceptance and the beam intensity loss. In these experiments, the electron-proton instability (electron cloud effect), which limits beam intensity in meson factories and in other large accelerators and storage rings [11], was first observed, explained, and suppressed [7,12]. In these experiments, the accumulated beam lives for 1 to 5 ms then the betatron oscillations grow and the beam is lost from the orbit over several dozen turns. A threshold intensity depends on beam intensity, electric field on electrodes surrounding beam, vacuum, Rf voltage and so all. At a high RF voltage is lost only central part of the beam. This instability was supressed by negative feed back from pickup beam position electrode 9 to deflector electrode 12 through an amplifier and phase shifter.

## PRODUCTION OF A CIRCULATING PROTON BEAM WITH COMPENSATED SPACE CHARGE AND COMPENSATION OF IONIZATION ENERGY LOSSES BY AN INDUCTION ELECTRIC FIELD.

Later in 1967, experiments were conducted on obtaining a circulating proton beam with compensated space charge and compensation of ionization energy losses by an induction electric field. The storage ring with betatron compensation of ionization losses is shown in Fig. 4. Between the poles of the electromagnet a hollow donut 3 was installed, into which a beam of neutrals was injected. The induction field was created by discharging the capacitor bank into a cut of a donut with a labyrinth, which prevented the magnetic field from penetrating into the donut. A selection of oscillogram traces characterizing the accumulation of a circulating proton beam in a donut is shown in Fig. 5 [13,14].

The circulating beam is accumulated to the equilibrium level shown in trace 1. The beam potential, measured by the ring pickup, increases and then decreases due to accumulation of compensating electrons within 10 μs. Then the potential rises sharply due to the ejection of electrons and the process of electron accumulation is repeated in trace 2. Collectors screened with a grid record the ejection of electrons and ions synchronously in traces 3, 4, 5, and 7. When the beam is bunched due to the effect of negative mass 8, the electrons do not accumulats and the instability is suppressed, shown in traces 6 and 9.

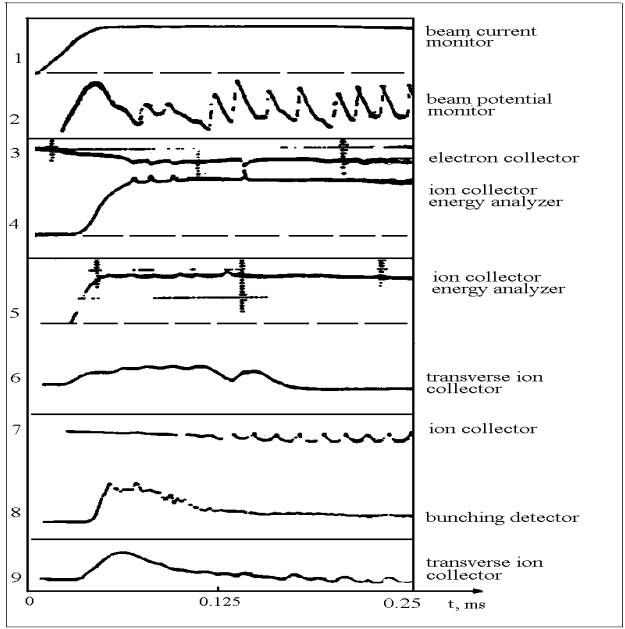

**FIGURE 5**: Selection of oscillogram traces that characterize the accumulation of a circulating proton beam in a donut in a quasi-betatron regime [13] of a circulating proton beam in a donut in a quasi-betatron regime [13]

To suppress the effect of negative mass, poles with strong focusing were added to the electromagnet. With these poles, the beam accumulation was studied with compensation of the ionization losses by the induction field. The beam accumulation signals with the compensation of the ionization losses by the induction field are shown in Fig. 6 a [11]. The beam current is accumulated and then saturates 1 with increasing signal from the horizontal loss probe 3. At this the signal from the vertical loss probe increases. 2.

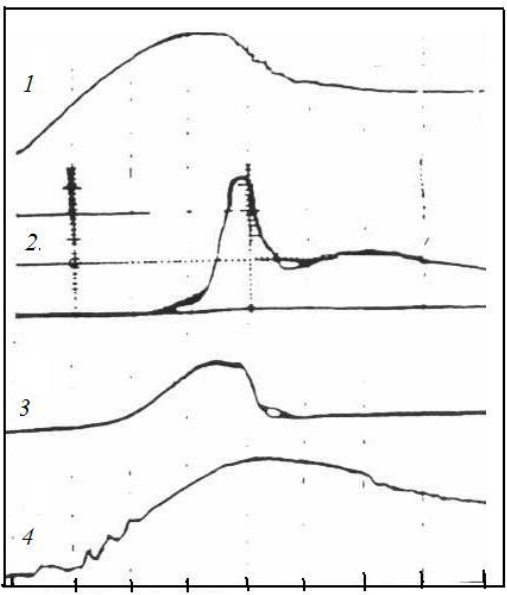
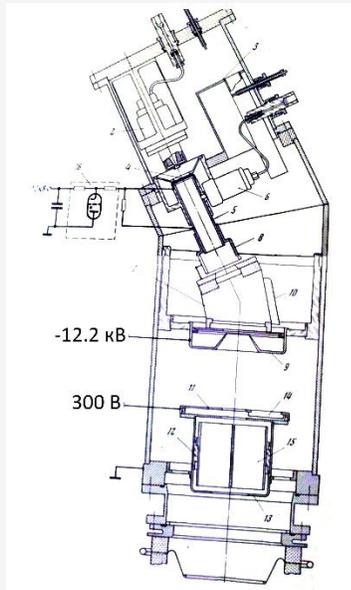

a                                                        b

**FIGURE 6**.a -Beam accumulation signals with strong focusing and compensation of ionization losses by an induction field [7].
1- Signal from the current meter, 2- signal from the vertical loss probe, 3- signal from the horizontal loss probe, 4- signal of the vertical coherent oscillations of the beam. The horizontal scale is 20 μs/div.
 b-schematic of a charge exchange hydrogen negative ion source for a Van de Graaf accelerator [19].

The pickup of the vertical position of the beam fixes the growth of vertical betatron oscillations 4 before the beam is lost vertically. This instability, associated with the oscillation of the compensating particles in the potential well of the beam, is well described by the theory of instability developed by B. V. Chirikov [15] for an electron beam compensated by ions (electron cloud effect). The study of collective effects in circulating beams with a maximum intensity with space charge compensation in combination with charge-exchange injection made it possible to create such a "super nonequilibrium" formation as a circulating proton beam compensated by an electron gas, with an intensity that is almost an order of magnitude larger than the spatial charge limit [16,17,18].

## PRODUCTION OF CIRCULATING PROTON BEAM WITH INTENSITY ABOVE THE SPACE CHARGE LIMIT

In 1970 a charge exchange H- ion source with current up to 20 mA was developed, as shown in Fig. 6 b [19]. This development helped to significantly increase accumulating current.

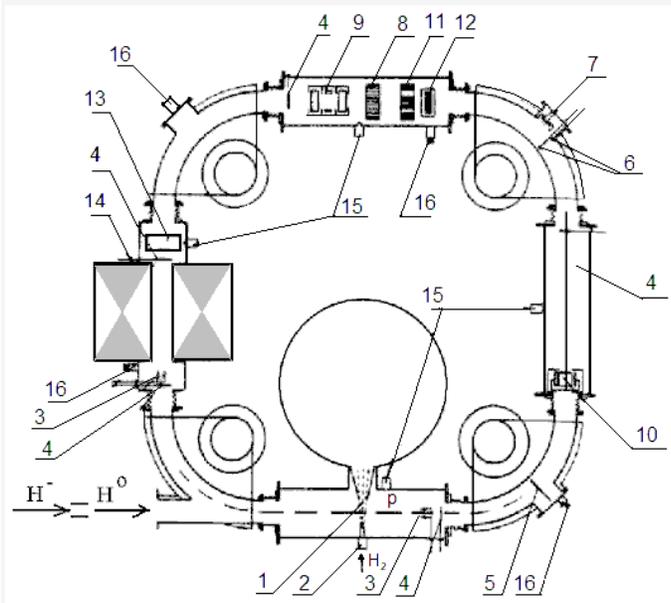

**FIGURE 7**. A schematic of storage ring for producing a circulating proton beam with an intensity above the space charge limit [15].
1- Supersonic jet-stripping target, 2-pulse jet valve, 3- beam collector, 4 quartz screen, 5, 6-mobile target, 7-ion collector, 8-Rogovsky belt, 9-beam position monitor, 10 Electrostatic pickup of quadrupole beam oscillations, 11- electromagnetic transverse beam oscillation sensor, 12-beam vertical beam sensor, 13-meter of secondary charged particles in the beam, 14-induction core, 15-pulse gas inlet, 16-stationary gas inlet.

The schematic of a storage ring for obtaining a circulating proton beam with intensity above the space charge limit is shown in Fig. 7. The oscillograms of the first observation of beam accumulation with an intensity exceeding the space charge limit for are shown in Fig. 8 [15]. The beam was accumulated to the space charge limit with the clearing of electrons along the entire orbit by electric field. This leads to the development of the Hereward instability, which does not lead to a loss of the beam. After turning off the electron clearing, the oscillations rapidly decay and the beam intensity in the orbit grows above the space charge limit [15] if the injection beam current is above the critical level (Fig. 8 a). If the injection beam current is below the critical level, after switching of the electron clearing voltage, the transverse oscillations grow and beam intensity is lost (Fig. 8 b). After increasing the injection current to 8 mA, it became possible to accumulate a beam with an intensity above the space charge limit without electron clearing. The process of accumulation is shown in Fig. 9. When the injection current is above the threshold, the current on the storage ring accumulates above the space charge limit (solid line). In this case, a lot of ions accumulate in the beam, and coherent oscillations rapidly decay. When the injection current is below critical, the current on the track is limited to a low level (dotted line), ions do not accumulate, coherent oscillations growing until the beam is lost. The results of the investigation of the charge-exchange method for the injection of protons into accelerators are presented in [20,21]. The inverse charge exchange of protons into atoms in targets from neutral particles limits the injection time at energies lower than $10^6$ eV. At higher energies, only multiple scattering of circulating protons is essential, limiting the injection time at a level of $10^4$ revolutions.

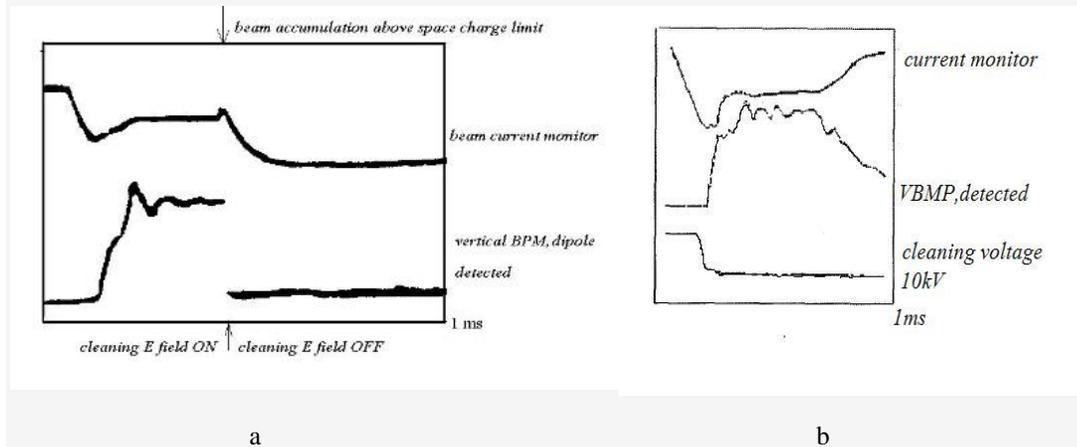

a  b

**FIGURE 8**. Oscillograms of the first observation of beam accumulation with an intensity exceeding the space charge limit. a-injection current above critical level; b- injection current below critical level [16].

By placing a target in the local minimum of the β-function of the focusing system, it is possible to reduce the effect of multiple scattering and even to ensure the attenuation of incoherent betatron oscillations due to ionization energy losses in the target. Such a large permissible duration of highly effective capture of protons allows us to sharply reduce the requirements for the intensity of the injected beam, and what is especially important is that with charge exchange injection, the requirements for the brightness of the injected beams are reduced. Compared to traditional injection methods that inject proton beams with the maximum allowable intensity and brightness, the same circulating proton beam intensities can be obtained by injecting H- beams with intensity and brightness a hundredfold less.

Sources available now and being developed do not provide injection into synchrotrons of sufficiently intense beams of ions polarized by nuclear spin. Sources with an intensity of $10^{-3}$ A have now been developed. By using charge-exchange injection, this intensity is enough to fill a booster synchrotron to the space charge limit [22]. Charge exchange injection can also be used for injection into cyclotrons. The injected particles can be passed to the center of the cyclotron in the form of neutral particles and stripped on a target located at the first turn. Such an injection method was used to inject polarized protons into in the cyclotrons.

## DEVELOPMENT OF CHARGE EXCHANGE INJECTION AROUND THE WORLD

In 1969, charge-exchange injection was successfully tested on the 12-GeV ZGS proton synchrotron in the USA at an injection energy of 50 MeV. Charge exchange injection of protons into a 200 MeV synchrotron was established as a prototype booster for the ZGS [23] and since 1977 served for many years as an intense pulsed neutron source (IPNS) [24,25]. In 1978, charge-exchange injection was established on the 8-GeV FNAL Booster at an injection energy of 200 MeV [26]. In 1982, the synchrotron AGS in BNL was transferred to the charge-exchange injection [27]. In 1984, charge-exchange injection was carried out in the ISIS synchrotron at RAL [28]. In 1980-84, charge exchange injection was established at KEK [29], at DESY [30] and at LANSCE in Los Alamos [31]. Charge-exchange injection is used on the CELSIUS storage ring (Uppsala, Sweden) [32], and in the COSY storage ring (Yulich Research Center, Germany) [33,34]. Charge-exchange injection was used in the synchrotron of the Institute of Experimental and Theoretical Physics for the accumulation of carbon ions [35]. Transition to charge-exchange injection is underway in the CERN booster [36] and in the IHEP booster [37]. The pulse intensity of a beam in a storage ring is limited by the development of the electron-proton instability (electron cloud effect). In 2006, the Spallation Neutron Source (SNS), an intense neutron source with charge-exchange injection in the Oak Ridge National Laboratory [38] was launched. Ions of H- with a current of 40 mA are accelerated in a superconducting linear accelerator to an energy of 1 GeV and accumulate during 1000 revolutions in a compact storage ring with a period of 1 μs rotation (accumulated current up to 50 A, power up to 50 GWt). Complex problems of the stability of the charge-exchange foil are discussed in [39]. In [40], laser ionization of H- ions into protons was tested. The history of the e-p instability detection is shown in Fig. 10 from [41]. In 1965, as is clear now, an e-p instability was observed in the ANL ZGS and the BNL AGS, but only at INP it was then correctly interpreted, explained and suppressed [41].

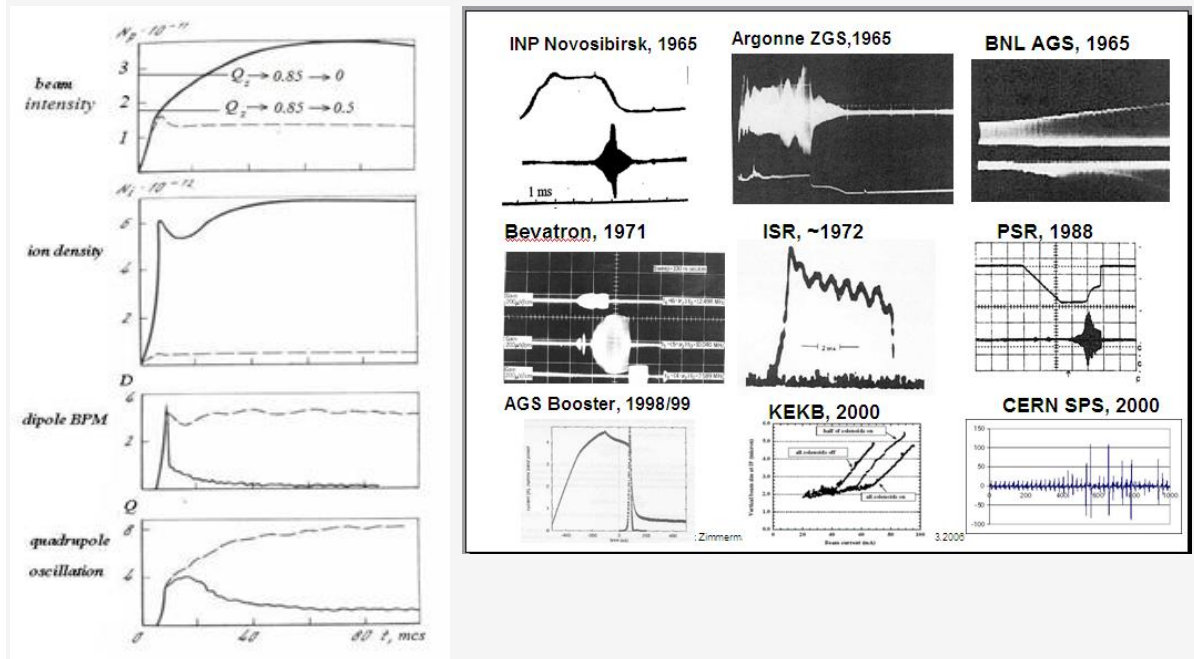

**FIGURE 9**. Proton beam accumulation with an intensity above space charge limit [17].

**FIGURE 10**. History of e-instability detection (electron cloud effect) from [41].

## ION SOURCES FOR CHARGE EXCHANGE INJECTION

In order to meet the requirements of change-exchange injection, H- ions with pulse intensity up to 100 mA and a high duty cycle are needed, which became possible after the discovery and development of a surface plasma method for producing the bright beams of negative ions in Institute of Nuclear Physics, Novosibirsk [42,43]. As example of ion source for charge exchange injection ken be a Penning discharge Surface plasma source developed in the Budker Institute of Nuclear Physics, Novosibirsk [44,45]. It can produce up to 150 ma of H- ions with frequency up to 400 Hz, pulse duration 0.25 ms. A review of negative ion sources for accelerators are given in publications [46,47,48]. Neutral beam injector for ITER with Surface Plasma Source of Negative Ion with cesiation, described in ITER Neutral Beam Test Facility - 2017

# REFERENCES

1 L.W. Alvarez, Rev. Sci. Instrum., 22, 705 (1951)
2 G.I. Dimov, The charge-exchange method for the injection of protons into accelerators and storage rings, Preprint 304, Institute of Nuclear Physics, Novosibirsk, 1969.
3 G.I. Budker, G.I. Dimov, Proceedings of the International Conference on Accelerators, Dubna, 1963. 933. M. 1964.
4 G.I. Dimov, V.G. Dudnikov, "Charge-exchange method for controlling particle beams", Sov. J. Plasma Phys. Engl. Transl. 4 (3) (1978).
5 G.I. Budker, G.I. Dimov, A.G. Popov, et al., Atomic Energy, 19, 507, 1965.
6 G. Budker, G. Dimov, V. Dudnikov, Proc. Int. Symp. on Electron and Positron Storage Ring, France, Sakley, 1966, rep. VIII, 6.1 (1966).
7 G. Budker, G. Dimov, V. Dudnikov, Sov. Atomic. Energy, 22, 348, (1967).
8 V. Dudnikov, Accumulation of an intense proton beam in a storage ring by the method of charge-exchange injection, Thesis for a candidate's degree of Phys.-Math. Nauk, INP SBRA USSR, Novosibirsk, 1966.
9 G. I. Dimov, I.Y. Timoshin, V.V. Demidov, V. G. Dudnikov, "Production beam of negative hydrogen ions with energy up to 1 MeV and current up to 1 mA", Pribory i technica experimenta, 4, 30 (1967).
10 GI Dimov, VG Dudnikov, ZhTF, 36, 1239 (1966).
11 G. Rumolo, A. Z. Ghalam, T. Katsouleas, et al., "Electron cloud effects on beam evolution in a circular accelerator", Physical Review Special Topics - Accelerators and Beams, 6, 081002 (2003).
12 GI Dimov, "Charge-exchange injection into accelerators and accumulators", doctoral dissertation, INP, 1968.
13 G. Budker, G. Dimov, V. Dudnikov, V. Shamovsky, "Experiments on electron compensation of proton beam in ring accelerator", Proc.VI Intern. Conf. On High energy accelerators, 1967, MIT & HU,A-104, CEAL-2000, (1967).
14 G. Dimov, V. Dudnikov, V. Shamovsky, Soviet Atomic Energy, 29, 5, 356 (1970).
15 BV Chirikov, Stability of a partially compensated electron beam, Atomic Energy, 36, 1239 (1966).
16 G. Dimov, V. Chupriyanov, V. Shamovsky. Sov. Phys., Tech. Phys., 16 (10), 1662 (1971).
17 Yu.I. Belchenko, G.I. Budker. G.I. Dimov, V.G. Dudnikov et al., Proceedings of the International Conference on Accelerators, Protvino, 1977; Preprint of INP 77-59, Novosibirsk, 1977.
18 G.I. Dimov, V.E. Chupriyanov, "Compensated proton-beam production in an accelerating ring", Particle Accelerators 14, 155-184 (1984).
19 G.I. Dimov, G. V. Roslyakov, " Injector of hydrogen negative ions with energy 1 Mev, current 20 mA, 0.2 ms", Pribory i technika experimenta, 2, 33 (1974).
20 G.I. Dimov, "Use of hydrogen negative ions in particle accelerators" Rev. Sci. Instrum. 67, 3393-3404 (1996).
21 M. Reiser, "Theory and Design of Charged Particle Beams", Wiley-VCH Verlag, second edition, (2009).
22 A.S. Belov, V.G. Dudnikov, V.E. Kuzik et al., Nucl. Instrum. Methods in Phys. Research, A333, 256 (1993).
23 J.D. Simpson, IEEE Trans. Nucl. Sci., NS-20, No. 3, 198 (19730.
24 R. Martin, Proc. VIII Internat. Conf. on High Energy accel. CERN, p. 540 (1971).
25 J. Simpson, R. Martin, R. Kustom, History of the ZGS 500 MeV booster, http://inspirehep.net/record/1322000
26 C. Hojvat, C. Ankenbrandt, B. Brown, et al., The Multiturn Charge Exchange Injection System for The Fermilab Booster Accelerator, IEEE Trans. on Nucl. Sci., NS-26, No. 3 (1979).
27 D.S. Barton, "Charge Exchange Injection at the AGS" BNL Int. Rep. 32784 PAC, Santa Fe NM, March 3 (1983).
28 V.C. Kempson, C.W. Planner and V.T. Pugh, Injection Dynamics and Multiturn Charge Exchange Injection into the Fast Cycling Synchrotron for the SNS, IEEE Trans. on Nucl. Sci., Vol. NS-28, No. 3, 3085 (1981).
29 T. Kawakubo, I. Sakai, H. Sasaki and M. Suetake, The H- Charge-Exchange Injection System in the Booster of the KEK 12 Gev Proton Synchrotron, http://inspirehep.net/record/234778/files/HEACC86_II_287 -290.pdf
30 L. Criegee, H. Dederichs, H. Ebel, G. Franke, D. M. Kong et al., The 50 MeV H− linear accelerator for HERA :LINAC3 collaboration, Rev. Sci. Instrum. 62, 867 (1991).
31 R. J. Macek et al, PAC 1993, p. 3739 (1993).
32 Hermanson L, et al., *in Proc. of workshop on Beam cooling and Related Topics* (Montreux, 4-8 October, 1993, CERN 94-03,26 April 1994), pp. 235
33 Baldin A M, Kovalenko A D *JINR Rapid Communications,* **377**.-96, Dubna,1996, p.5
34 Sidorin A.O. "Forming intense ion beams in storage rings with multiturn charge-exchange injection and electron cooling ", Dissertation (Dubna: 2003)
35 Alekseev N.N., Koshkarev D G, Sharkov B Yu *JETP Letters* **77**, 123 (2003)
36 J. Lettry, D. Aguglia, J. Alessi, et al., CERN's Linac4 H− sources: Status and operational results, AIP Conference Proceedings 1655, 030005 (2015).
37 B.A. Frolov, V.S. Klenov, V. N. Mihailov, Simulation and Optimization Of Ion Optical Extraction, Acceleration and H- Ion Beam Matching Systems**,** Proceedings of RuPAC2014, Obninsk, Kaluga Region, Russia, 2014.
38 J. Wei et al, PRST-AB, 3, 080101(1999).
39 M. Plum, AAC - HEBT/Ring/RTBT Overview, Plum-AAC-Feb10_r1.pptx 2010.

40 Sarah Cousineau,† Abdurahim Rakhman, Martin Kay, A. Aleksandrov, et al., First Demonstration of Laser-Assisted Charge Exchange for Microsecond Duration H− Beams, PRL 118, 074801 (2017).
41 F. Zimmermann, Review of single bunch instability driven by electron cloud", Phys. Rev. ST AB, 7, 124801 (2004).
42 V. Dudnikov, SU Author sertificate, Method of negative ion production., 411542, filed 10 / III, 1972, Published on 15 / 1.1974, Bulletin No.2. http://www.findpatent.ru/patent/41/411542.html
43 Yu Belchenko, G. Dimov, V. Dudnikov, Physical principles of surface plasma sources operation, Symposium on the Production and Neutralization of Negative Hydrogen Ions and Beams, Brookhaven, 1977 (Brookhaven National Laboratory (BNL), Upton, NY, 1977) pp. 79-96;
44 VG Dudnikov, A source of negative hydrogen ions with Penning geometry, Proceedings of the IV All-Union Conference on Accelerators of Charged Particles, M. Nauka, 1975, vol. 1, p. 323.
45 G.I. Dimov, G.E. Derevyankin, V.G. Dudnikov, IEEE Trans. Nucl. Sci., NS24, No3 (1977).
46 M. Stockli, "Pulsed, high-current H- Ion Sources for Future Accelerators", 145, ICFA Beam Dynamics Newsletter No. 73, Issue Editor: G. Machicoane and P. N. Ostroumov , Editor in Chief: Y. H. Chin (1918).
47 V. Dudnikov, "Modern high intensity H- accelerator sources", http://arxiv.org/abs/1806.03391
48 Dan Faircloth and Scott Lawrie, "An overview of negative hydrogen ion sources for accelerators", New Journal of Physics, 20, (2018).  http://iopscience.iop.org/article/10.1088/1367-2630/aaa39e/pdf